\documentclass[reprint,twocolumn,prl,superscriptaddress,showpacs,showkeys,aps]{revtex4-2}

\usepackage{epsfig,amsmath,amsfonts,amssymb}
\usepackage{amsmath,amsfonts,amssymb}
\usepackage{graphicx}
\usepackage{subfigure}
\usepackage{mathbbol}
\usepackage{amsthm}

\begin{document}

\title{Ultimate conversion efficiency bound for the forward double-$\Lambda$ atom-light coupling scheme}

\author{Dionisis Stefanatos}
\email{dionisis@post.harvard.edu}
\affiliation{Materials Science Department, School of Natural Sciences, University of Patras, Patras 26504, Greece}
\author{Athanasios Smponias}
\affiliation{Materials Science Department, School of Natural Sciences, University of Patras, Patras 26504, Greece}
\author{Hamid Reza Hamedi}
\affiliation{Institute of Theoretical Physics and Astronomy, Vilnius University, Saul\.{e}tekio 3, Vilnius LT-10257, Lithuania}
\author{Emmanuel Paspalakis}%
\affiliation{Materials Science Department, School of Natural Sciences, University of Patras, Patras 26504, Greece}

\date{\today}

\begin{abstract}
We show that for the two widely used configurations of the double-$\Lambda$ atom-light coupling scheme, one where the control fields are applied in the same $\Lambda$-subsystem and another where they applied in different $\Lambda$-subsystems, the forward propagation of the probe and signal fields is described by the same set of equations. We then use optimal control theory to find the spatially-dependent optimal control fields which maximize the conversion efficiency from the probe to the signal field, for a given optical density. The present work is expected to find application in the implementation of efficient frequency and orbital angular momentum conversion devices for quantum information processing, as well as to be useful to many other applications using the double-$\Lambda$ atom-light coupling scheme.
\end{abstract}

\maketitle

\begin{figure}[t]
 \centering
 \fbox{
		\begin{tabular}{cc}
       \subfigure[Direct configuration]{
	            \label{fig:direct}
	            \includegraphics[width=0.43\linewidth]{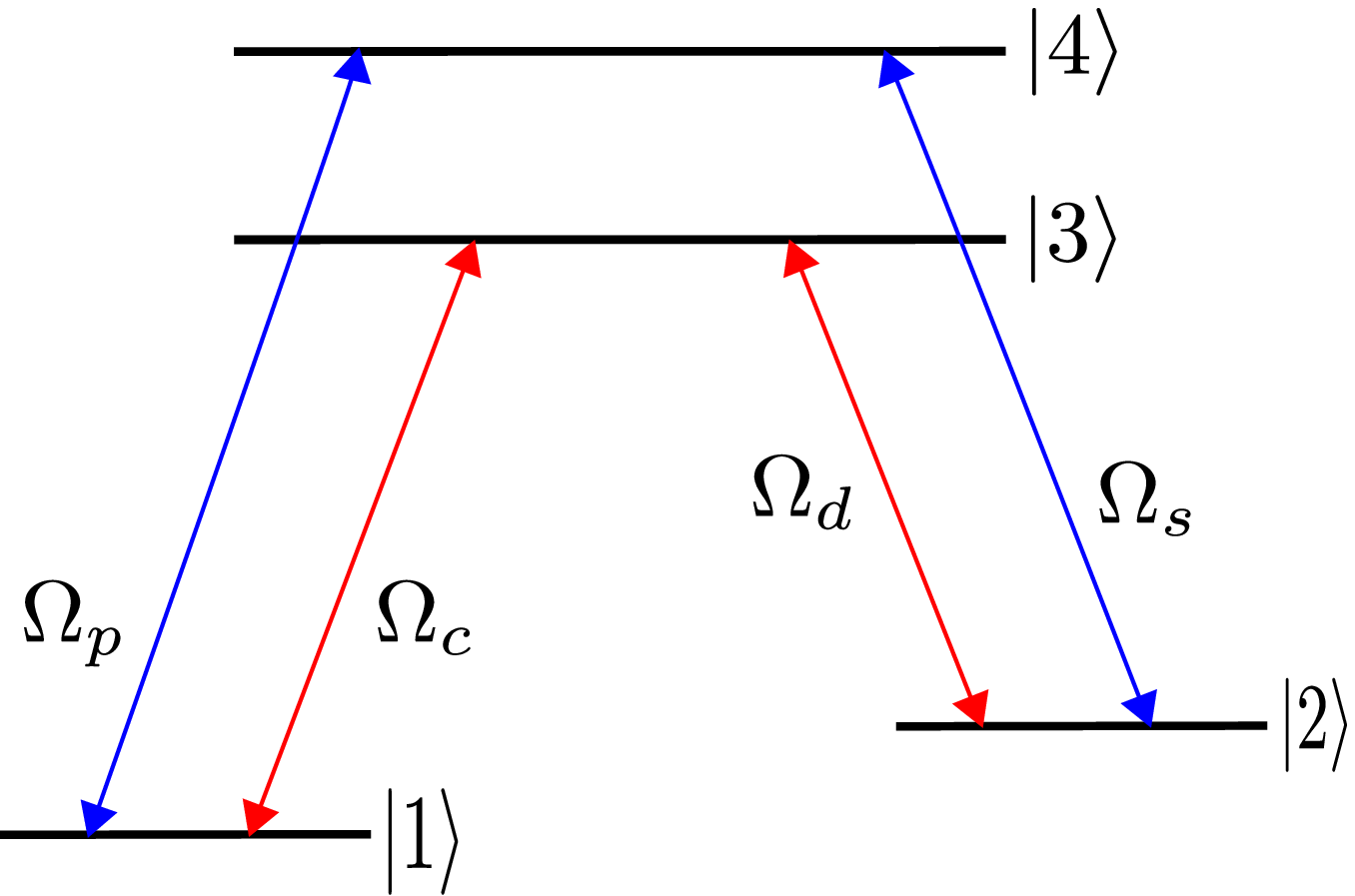}} &
       \subfigure[Mixed configuration]{
	            \label{fig:mixed}
	            \includegraphics[width=0.43\linewidth]{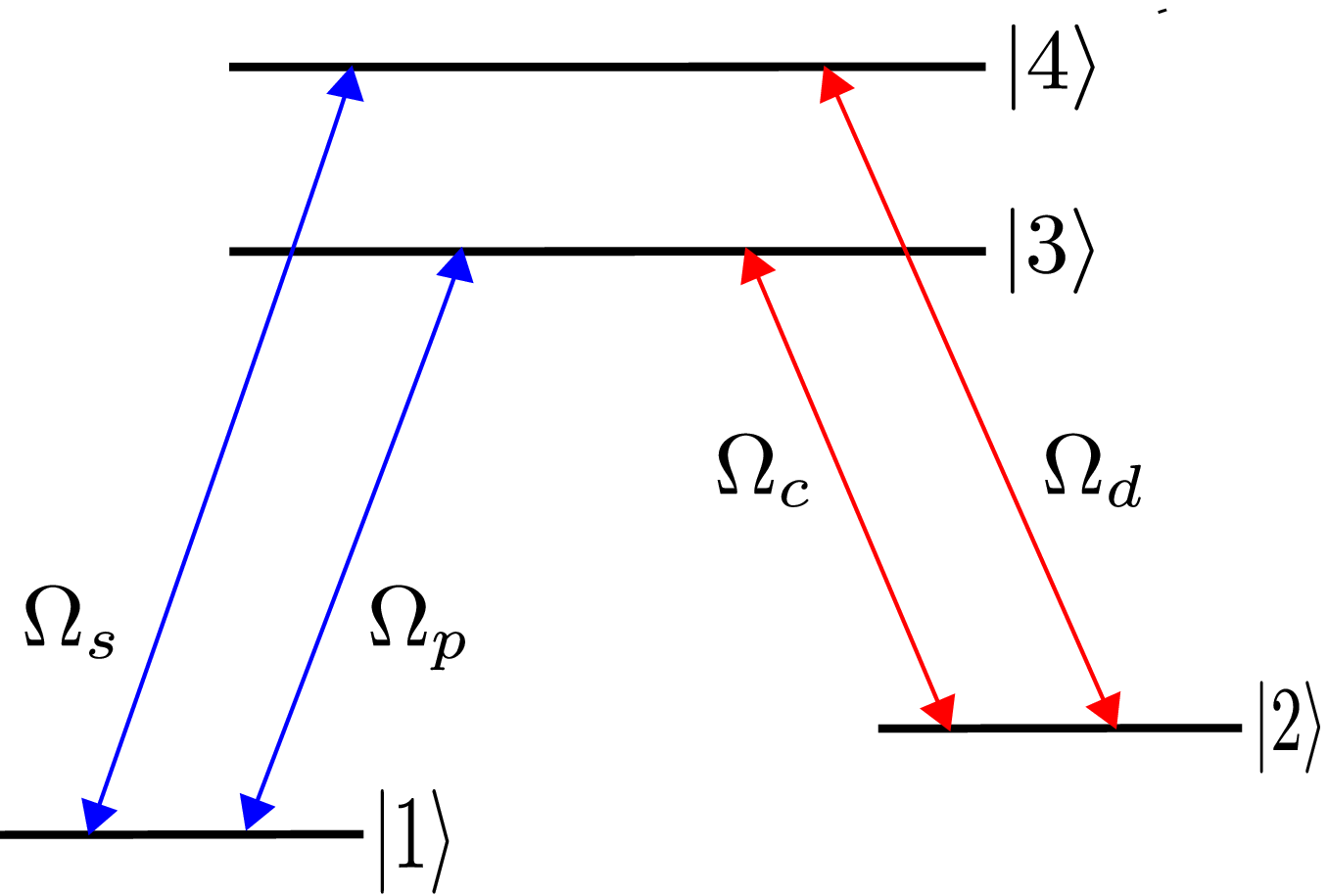}}
		\end{tabular}
 }
\caption{(a) In the direct configuration, the probe and signal fields $\Omega_p, \Omega_s$ appear in the same $\Lambda$-subsystem, while the control fields $\Omega_c, \Omega_d$ are applied in the other. (b) In the mixed configuration, the probe and signal fields appear in different $\Lambda$-subsystems. This configuration can also be viewed as built by a $\Lambda$- and a $V$-system \cite{Kocharovskaya90}.}
\label{fig:configuration}
\end{figure}

During the last few decades, the double-$\Lambda$ atom-light coupling scheme \cite{Lukin00} has received considerable attention among the quantum optics community. This system is a four-state unbranched loop where the labeling of the four applied laser fields can be considered as two linked $\Lambda$-subsystems, as shown in Fig. \ref{fig:configuration}. Usually, a pair of strong control fields $\Omega_c, \Omega_d$, which can be spatially-dependent, prepare atoms in a coherent superposition state. This coherently prepared atomic cloud can be used to manipulate the properties of the weak probe and signal fields $\Omega_p, \Omega_s$ propagating inside it. For example, when the signal field is absent at the entrance to the medium, this setup can generate it by converting the probe field.

The double-$\Lambda$ atom-light coupling scheme finds a wide range of applications, including light storage \cite{Raczynski04,Phillips11}, generation of squeezed light states \cite{McCormick07,Turnbull13}, phase-controlled light switching \cite{Kang06,Xu13}, frequency conversion \cite{Lee16,Liu17,Merriam00,Moiseev06,Juo18,Kang04,Wang10,Chiu14,Eilam06} and orbital angular momentum conversion \cite{Hamedi18a,Hamedi19,Prajapati19} between light beams, as well as many others \cite{Kocharovskaya90,Deng05,Eilam08,Hsiao14,Liu16,Jeong16,Zhang15}. In most of these works, two configurations of the applied laser fields are usually encountered. In the first, which we call the direct configuration and display it in Fig. \ref{fig:direct}, the probe and signal fields appear in the same $\Lambda$-subsystem, while the control fields are applied in the other. In the other configuration, shown in Fig. \ref{fig:mixed}, the probe and signal fields appear in different $\Lambda$-subsystems; we call this the mixed configuration.

In the present work we show that, under some reasonable approximations, the forward propagation of the probe and signal fields in both configurations is described by the same set of equations. Then, we use optimal control theory \cite{Bryson} to find the optimal control fields $\Omega_c, \Omega_d$ which maximize the conversion efficiency from $\Omega_p$ to $\Omega_s$ for a given optical density. We emphasize that we concentrate on the forward scheme, where $\Omega_p, \Omega_s$ propagate in the same direction, which suffers less from phase-mismatch than the backward propagation scheme \cite{Juo18}. Also, we consider the case where all fields are resonant with the corresponding transitions, so the interaction strength between atoms and photons is maximized.

We consider first the mixed configuration, shown in Fig. \ref{fig:mixed}. The Maxwell-Bloch equations governing the dynamics of the fields $\Omega_p,\Omega_s$ and the atomic coherences $\rho_{21},\rho_{31},\rho_{41}$ are \cite{Juo18}
\begin{subequations}
\label{rho}
\begin{eqnarray}
\frac{d}{dt}\rho_{31}&=&i(\Omega_p+\Omega_c\rho_{21})-\Gamma_{31}\rho_{31}\\
\frac{d}{dt}\rho_{41}&=&i(\Omega_s+\Omega_d\rho_{21})-\Gamma_{41}\rho_{41}\\
\frac{d}{dt}\rho_{21}&=&i(\Omega^*_c\rho_{31}+\Omega^*_d\rho_{41})-\Gamma_{21}\rho_{21}
\end{eqnarray}
\end{subequations}
and
\begin{subequations}
\label{maxwell}
\begin{eqnarray}
\frac{\partial}{\partial z}\Omega_p+\frac{1}{c}\frac{\partial}{\partial t}\Omega_p&=&i\frac{\alpha_p\Gamma_{31}}{2L}\rho_{31},\\
\frac{\partial}{\partial z}\Omega_s+\frac{1}{c}\frac{\partial}{\partial t}\Omega_s&=&i\frac{\alpha_s\Gamma_{41}}{2L}\rho_{41},
\end{eqnarray}
\end{subequations}
where $\Gamma_{31}, \Gamma_{41}$ are the spontaneous decay rates of the excited states, $\Gamma_{21}$ is the dephasing rate of the ground states, $\alpha_p, \alpha_s$ are the optical densities of the probe and signal transitions, and $L$ is the length of the medium. To simplify the model, we assume
$\Gamma_{31}=\Gamma_{41}=\Gamma$, $\Gamma_{21}=0$, $\alpha_p=\alpha_s=\alpha$, justified in Refs.\ \cite{Juo18,Hamedi19}. Note also that in Eq. (\ref{rho}) we use the approximation $\rho_{11}\approx 1$, which is valid to first order for weak probe and signal fields.
If we solve for the steady state of system (\ref{rho}), we get to the first order (see the supplemental document for more details)
\begin{equation}
\label{rho_omega}
\left(
\begin{array}{c}
  \rho_{31}\\
  \rho_{41}
\end{array}
\right)
=
\frac{i}{\Gamma}
\left(
\begin{array}{cc}
  \cos^2{\theta} & -\sin{\theta}\cos{\theta}\\
  -\sin{\theta}\cos{\theta} & \sin^2{\theta}
\end{array}
\right)
\left(
\begin{array}{c}
  \Omega_p\\
  \Omega_s
\end{array}
\right),
\end{equation}
where $\theta$ is the mixing angle defined by the equation
\begin{equation}
\label{mixing_angle}
\Omega_c=\Omega\sin{\theta},\quad\Omega_d=\Omega\cos{\theta}
\end{equation}
and $\Omega$ being the generalized Rabi frequency.
Substituting Eq. (\ref{rho_omega}) into Eq. (\ref{maxwell}) we obtain in the steady state the following coupled equations for the propagation of the probe and signal pulses
\begin{equation}
\label{omega}
\frac{\partial}{\partial z}
\left(
\begin{array}{c}
\Omega_p\\
  \Omega_s
\end{array}
\right)
=
-\frac{\alpha}{2L}
\left(
\begin{array}{cc}
  \cos^2{\theta} & -\sin{\theta}\cos{\theta}\\
  -\sin{\theta}\cos{\theta} & \sin^2{\theta}
\end{array}
\right)
\left(
\begin{array}{c}
  \Omega_p\\
  \Omega_s
\end{array}
\right).
\end{equation}
The above system, which describes pulse propagation for the mixed configuration, is exactly the same with the corresponding system for the direct configuration, derived in Ref.\ \cite{Hamedi19}. If we normalize the propagation distance $z$ with the absorption length $L_{abs}=L/\alpha$ and use the dimensionless variable $\zeta=z/L_{abs}$, we get
\begin{equation}
\label{system}
\left(
\begin{array}{c}
  \dot{\Omega}_p\\
  \dot{\Omega}_s
\end{array}
\right)
=
-\frac{1}{2}
\left(
\begin{array}{cc}
  \cos^2{\theta} & -\sin{\theta}\cos{\theta}\\
  -\sin{\theta}\cos{\theta} & \sin^2{\theta}
\end{array}
\right)
\left(
\begin{array}{c}
  \Omega_p\\
  \Omega_s
\end{array}
\right),
\end{equation}
where the dot denotes the derivative $\partial/\partial\zeta$. Note that, under this transformation, the propagation takes place from $\zeta=0$ to $\zeta=\alpha$.

In order to study pulse propagation under a position-dependent mixing angle $\theta(\zeta)$, i.e. spatially varying control fields $\Omega_c(\zeta),\Omega_d(\zeta)$, it is more convenient to work in the adiabatic frame. The eigenstates of the propagation matrix in Eq. (\ref{system}) are
\begin{equation}
\label{eigenstates}
\psi_0=\left(
\begin{array}{c}
  \sin{\theta}\\
  \cos{\theta}
\end{array}
\right),
\quad
\psi_{-1/2}=\left(
\begin{array}{c}
  \cos{\theta}\\
  -\sin{\theta}
\end{array}
\right),
\end{equation}
with corresponding eigenvalues $0$ and $-1/2$, respectively.
The transformation to the adiabatic basis is
\begin{equation}
\label{transformation}
\left(
\begin{array}{c}
  y\\
  x
\end{array}
\right)
=
\left(
\begin{array}{cc}
  \sin{\theta} & \cos{\theta}\\
  \cos{\theta} & -\sin{\theta}
\end{array}
\right)
\left(
\begin{array}{c}
  \Omega_p\\
  \Omega_s
\end{array}
\right).
\end{equation}
Using Eqs. (\ref{system}) and (\ref{transformation}) we find
\begin{equation}
\label{adiabatic}
\left(
\begin{array}{c}
  \dot{y}\\
  \dot{x}
\end{array}
\right)
=
\left(
\begin{array}{cc}
  0 & -u\\
  u & -\frac{1}{2}
\end{array}
\right)
\left(
\begin{array}{c}
  y\\
  x
\end{array}
\right),
\end{equation}
where the control function $u(\zeta)$ is defined as
\begin{equation}
\label{theta}
\dot{\theta}=-u.
\end{equation}
For $\Omega_p(0)=\Omega_0, \Omega_s(0)=0$ and the boundary conditions
\begin{equation}
\label{b_theta}
\theta(0)=\frac{\pi}{2},\quad\theta(\alpha)=0
\end{equation}
we get
\begin{equation}
\label{initial}
x(0)=0,\quad y(0)=\Omega_p(0)=\Omega_0
\end{equation}
and
\begin{equation}
\label{final}
x(\alpha)=\Omega_p(\alpha),\quad y(\alpha)=\Omega_s(\alpha).
\end{equation}
It is worth to point out that for large $\alpha$, if $\theta$ varies slowly between the values in Eq. (\ref{b_theta}), with rate $u=|\dot{\theta}|\ll 1$, then $y(\zeta)$ remains constant during the evolution, thus $\Omega_s(\alpha)=y(\alpha)\approx y(0)=\Omega_p(0)=\Omega_0$. The slow conversion from the initial $\Omega_p$ to the final $\Omega_s$ takes place along the eigenstate $\psi_0$ of the original system (\ref{system}).

We next move to find the optimal control $u(\zeta)$, $0\leq\zeta\leq \alpha$, for given finite optical density $\alpha$, which maximizes the final value $\Omega_s(\alpha)=y(\alpha)$.
The control Hamiltonian is \cite{Bryson}
\begin{equation}
\label{Hc}
H_c=\lambda_x\dot{x}+\lambda_y\dot{y}+\mu u=(\lambda_x y-\lambda_y x+\mu)u-\frac{1}{2}\lambda_x x,
\end{equation}
where $\lambda_x,\lambda_y,\mu$ are the Lagrange multipliers corresponding to $x,y,\theta$. They satisfy the adjoint equations
\begin{subequations}
\label{lambda}
\begin{eqnarray}
\dot{\lambda}_y&=-\frac{\partial H_c}{\partial y}&=-u\lambda_x,\label{ly}\\
\dot{\lambda}_x&=-\frac{\partial H_c}{\partial x}&=u\lambda_y+\frac{1}{2}\lambda_x,\label{lx}
\end{eqnarray}
\end{subequations}
while $\mu$ is constant since $\theta$ is a cyclic variable. The optimal control $u(\zeta)$ is chosen to maximize the control Hamiltonian $H_c$ \cite{Bryson}, which is a mathematical construction, so its maximization does not correspond to maximizing the energy of some system but the target quantity $y(\alpha)$. Note that we have not imposed any bound on $u$, so Eqs. (\ref{adiabatic}) and (\ref{theta}) are fully equivalent to the original system (\ref{system}). Even infinite values are allowed momentarily, corresponding to instantaneous jumps in the angle $\theta$. Since $H_c$ is a linear function of $u$ with coefficient $\phi=\lambda_x y-\lambda_y x+\mu$, if $\phi\neq 0$ for a finite interval then the corresponding optimal control should be $\pm\infty$ for the whole interval, which is obviously unphysical. We conclude that $\phi=0$ almost everywhere, except some isolated points where jumps in the angle $\theta$ can occur. The optimal control which maintains this condition is called \emph{singular} \cite{Bryson}. Such controls have been exploited in nuclear magnetic resonance to minimize the effect of relaxation \cite{Lapert10}. In order to find the singular optimal control $u_s$ we additionally use the conditions $\dot{\phi}=\ddot{\phi}=0$ and obtain the following equations
\begin{subequations}
\label{phi}
\begin{eqnarray}
\lambda_y x-\lambda_x y&=&\mu,\label{p}\\
\lambda_y x+\lambda_x y&=&0,\label{dp}\\
2(\lambda_y y-\lambda_x x)u_s&=&\frac{1}{2}(\lambda_y x-\lambda_x y).\label{ddp}
\end{eqnarray}
\end{subequations}
Solving for $\lambda_x,\lambda_y,u_s$ we find
\begin{equation}
\label{lxy}
\lambda_x=-\frac{\mu}{2y},\quad \lambda_y=\frac{\mu}{2x},
\end{equation}
and
\begin{equation}
\label{feedback}
u_s=\frac{xy}{2(x^2+y^2)}.
\end{equation}

\begin{figure}[t]
\centering
\fbox{\includegraphics[width=0.6\linewidth]{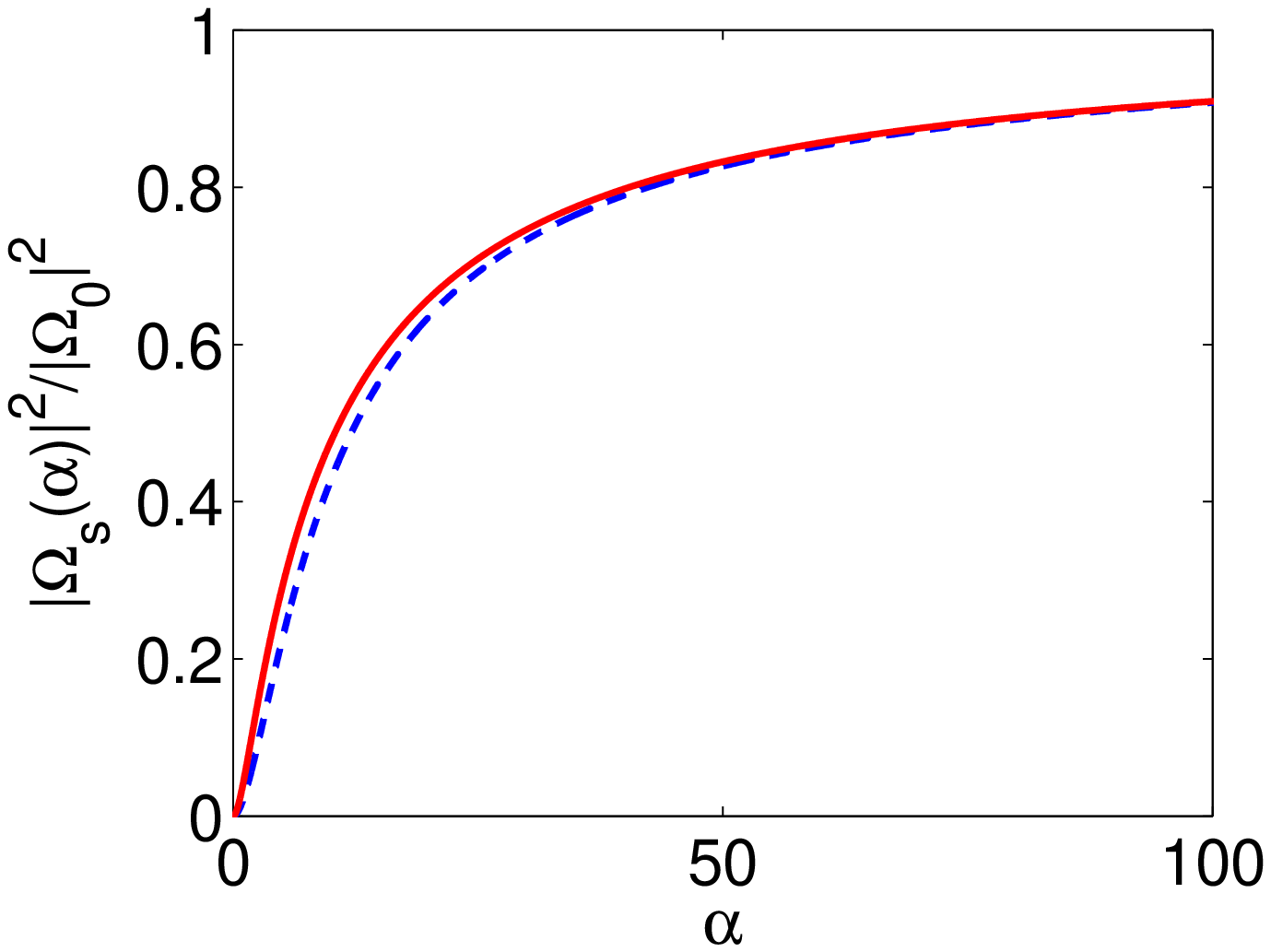}}
\caption{Conversion efficiency for the optimal protocol (red solid line) and for the constant control protocol of Eq.\ (\ref{constant_u}) (blue dashed line).}
\label{fig:efficiency}
\end{figure}

\begin{figure}[t]
 \centering
 \fbox{
		\begin{tabular}{cc}
       \subfigure[Control fields]{
	            \label{fig:controls}
	            \includegraphics[width=0.43\linewidth]{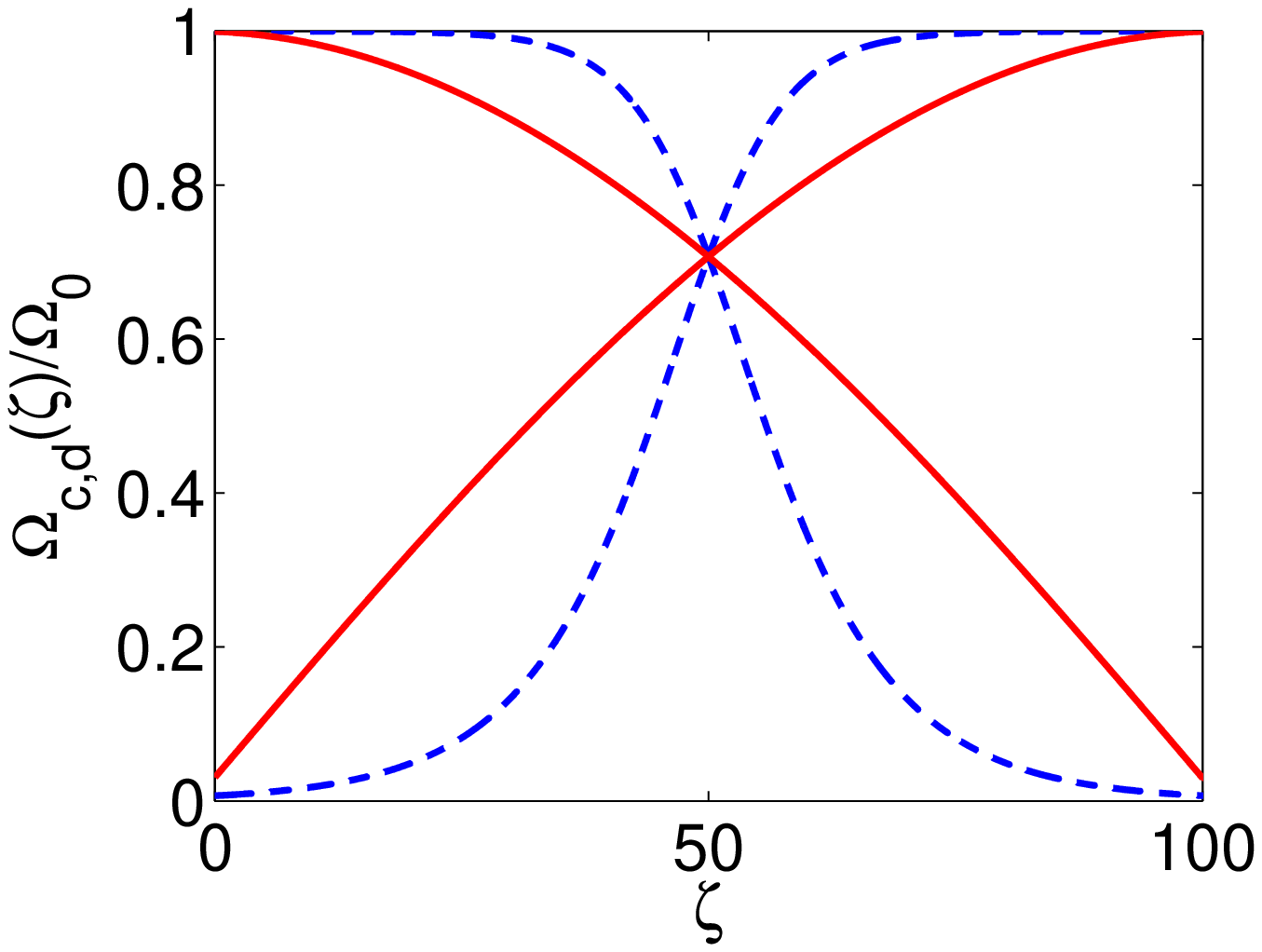}} &
       \subfigure[Intensities of probe and signal fields]{
	            \label{fig:prop_theory}
	            \includegraphics[width=0.43\linewidth]{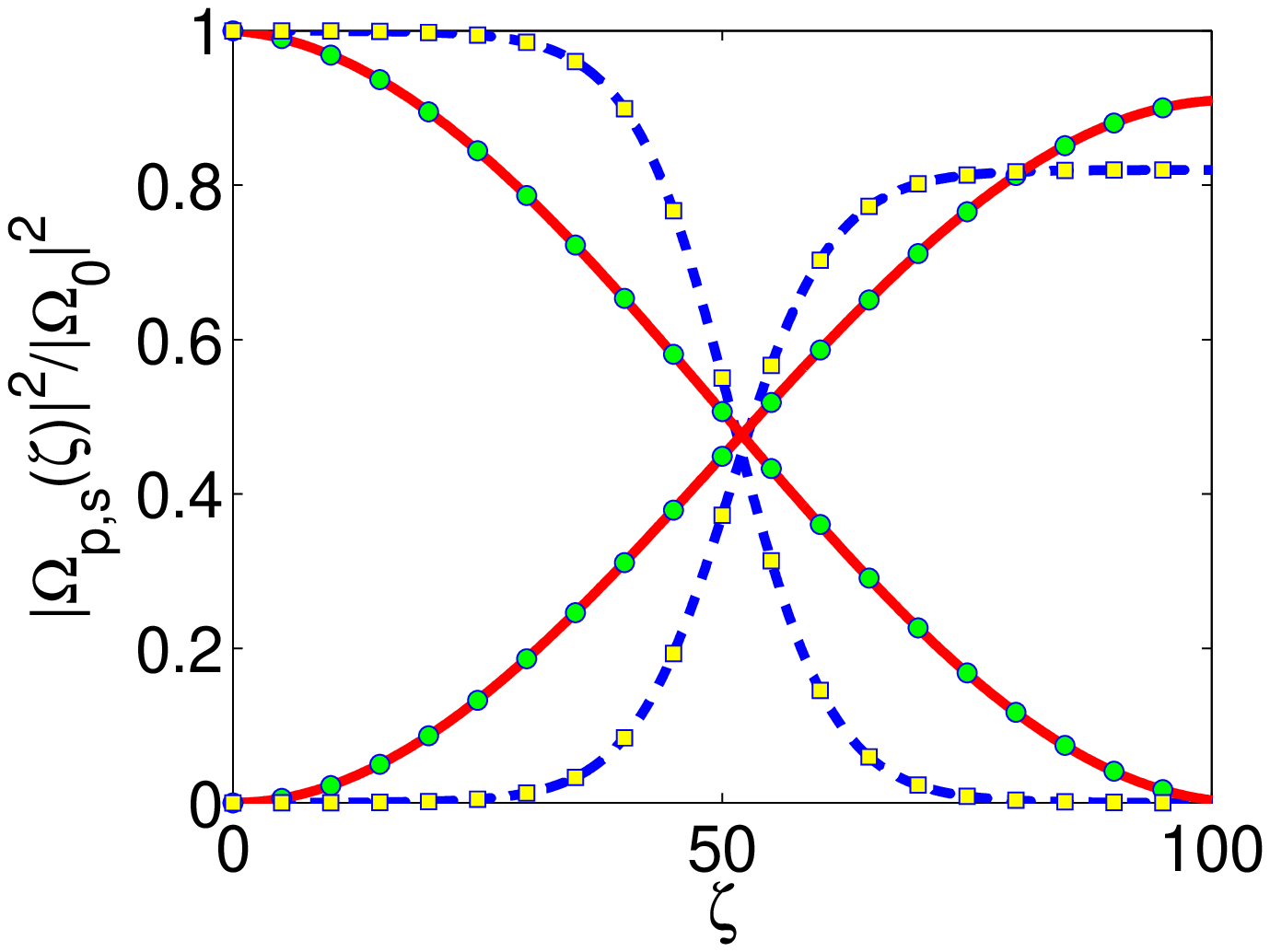}}
		\end{tabular}
 }
\caption{(a) Dimensionless control fields for the optimal protocol (red solid lines) and the adiabatic protocol (\ref{adiabatic_controls}) (blue dashed lines), as functions of the normalized distance $\zeta=z/L_{abs}$, for optical density $\alpha=100$. (b) Propagation of the corresponding probe and signal pulses using the approximate Eq. (\ref{system}) (red solid and blue dashed lines) and the full Maxwell-Bloch equations (green circles and yellow squares).}
\label{fig:example}
\end{figure}

According to optimal control theory \cite{Bryson}, the control Hamiltonian for a system without explicit dependence on the running variable $\zeta$, as in our case, is constant. Using $\phi=0$ and the expression (\ref{lxy}) for $\lambda_x$  in Eq. (\ref{Hc}), we conclude that the singular arc is a straight line passing through the origin of the $xy$-plane,
\begin{equation}
\label{slope}
\frac{y}{x}=\mbox{const.}=\tan{\theta_0}.
\end{equation}
This implies that the singular control $u_s$, given in Eq. (\ref{feedback}) in terms of a feedback law, is also constant. Now we can describe the optimal pulse-sequence. There is a delta pulse at $\zeta=0$, resulting in a jump from $\theta(0^-)=\pi/2$ to $\theta(0^+)=\theta_0$, where the initial angle is to be determined. The system is brought on the singular arc and remains there for $0<\zeta<\alpha$. During this interval the angle decreases linearly with slope $u_s$, $\theta(\zeta)=\theta_0-u_s\zeta$. At the final distance $\zeta=\alpha$, another delta pulse changes the angle from
$\theta(\alpha^-)=\theta_0-u_s\alpha$ to $\theta(\alpha^+)=0$. The optimal control has the form \emph{bang-singular-bang},
\begin{equation}
\label{pulse_sequence}
u(\zeta)=\left\{\begin{array}{cl} (\pi/2-\theta_0)\delta(\zeta), & \zeta=0 \\u_s, & 0<\zeta<\alpha\\ (\theta_0-u_s\alpha)\delta(\zeta-\alpha), & \zeta=a \end{array}\right..
\end{equation}
Observe from the original system (\ref{system}) that jumps in the angle $\theta$ do not change $\Omega_p,\Omega_s$. In the adiabatic basis these jumps are accompanied by sudden rotations of the $(x,y)^T$ vector such that $\Omega_p,\Omega_s$ remain unchanged, see the transformation (\ref{transformation}). Thus, in order to implement the optimal protocol, one simply needs to vary linearly the angle from $\theta_0$ to $\theta_0-u_s\alpha$ with the slope $u_s<\pi/(2\alpha)$. 

We now find the optimal $\theta_0$ corresponding to a given optical density $\alpha$. After the application of the pulse sequence (\ref{pulse_sequence}) to system (\ref{adiabatic}) we get
\begin{equation}
\label{max_eff}
\frac{|\Omega_s(\alpha)|^2}{|\Omega_0|^2}=e^{-2\gamma\alpha}\sin^2{(u_s\alpha)},
\end{equation}
where
\begin{equation}
\label{u_theta}
u_s=\frac{\tan{\theta_0}}{2(1+\tan^2{\theta_0})}=\frac{1}{4}\sin{(2\theta_0)}
\end{equation}
from Eqs. (\ref{feedback}), (\ref{slope}), and
\begin{equation*}
\gamma=\frac{1}{2(1+\tan^2{\theta_0})}.
\end{equation*}
If we optimize the rhs of Eq. (\ref{max_eff}) with respect to $\theta_0$ we obtain
\begin{equation}
\label{preliminary}
\tan{(u_s\alpha)}=\frac{du_s}{d\theta_0}/\frac{d\gamma}{d\theta_0}=\frac{\tan^2{\theta_0}-1}{2\tan{\theta_0}}=\tan{\left(2\theta_0-\frac{\pi}{2}\right)}.
\end{equation}
Since $0<u_s\alpha<\pi/2$ and $\pi/4<\theta_0<\pi/2$, where the lower bound in the latter relation comes from the third term in Eq. (\ref{preliminary}), from the monotonicity of the $\tan$ function in $(0,\pi/2)$ we conclude that
\begin{equation}
\label{transcendental}
\frac{\alpha}{4}\sin{(2\theta_0)}=2\theta_0-\frac{\pi}{2},
\end{equation}
where we have also used Eq. (\ref{u_theta}). For a given optical density $\alpha$, this is a transcendental equation for $\theta_0$. It can be easily proved that it has a unique solution in the interval $\pi/4<\theta_0<\pi/2$. Note that for $\alpha\rightarrow 0$ it is $\theta_0\rightarrow\pi/4$, while for $\alpha\rightarrow \infty$ it is $\theta_0\rightarrow\pi/2$. The limiting values of the optimal conversion efficiency are
\begin{equation}
\label{limits_opt}
\frac{|\Omega_s(\alpha)|^2}{|\Omega_0|^2}=\left\{\begin{array}{cl} \frac{1}{16}\alpha^2, & \alpha\ll 1 \\1-\frac{\pi^2}{\alpha}, & \alpha\gg 1\end{array}\right.,
\end{equation}
thus for sufficiently large optical density the conversion efficiency approaches unity. In Fig. \ref{fig:efficiency} we plot the optimal conversion efficiency (red solid line) as a function of the optical density, for $0\leq\alpha\leq 100$.

As we pointed out above, for large $\alpha$ we have $\theta_0\rightarrow\pi/2$, thus the boundary jumps become smaller. In this limit, the optimal solution tends to a linear decrease of the angle $\theta$ from $\pi/2$ to $0$,
\begin{equation}
\label{constant_u}
u(\zeta)=\mbox{const.}=\frac{\pi}{2\alpha},\quad\theta(\zeta)=\frac{\pi}{2}-u\zeta.
\end{equation}
Under this constant control protocol, ensuring that $\Omega_c(0)=\Omega_d(\alpha)=\Omega_0$ and $\Omega_c(\alpha)=\Omega_d(0)=0$, Eq.\ (\ref{adiabatic}) can be easily integrated and at the final distance $\zeta=\alpha$ one finds \cite{Lee16,Juo18}
\begin{equation}
\label{constant_eff}
\frac{|\Omega_s(\alpha)|^2}{|\Omega_0|^2}=e^{-\eta\alpha}\left[\cosh{(\kappa\alpha)}+\frac{\eta}{2\kappa}\sinh{(\kappa\alpha)}\right]^2,
\end{equation}
where
\begin{equation*}
\eta=\frac{1}{2},\quad\kappa=\sqrt{\left(\frac{\eta}{2}\right)^2-u^2}.
\end{equation*}
From Eq. (\ref{constant_eff}) we find the limiting cases
\begin{equation}
\label{limits_const}
\frac{|\Omega_s(\alpha)|^2}{|\Omega_0|^2}=\left\{\begin{array}{cl} \frac{1}{4\pi^2}\alpha^2, & \alpha\ll 1 \\1-\frac{\pi^2}{\alpha}, & \alpha\gg 1\end{array}\right..
\end{equation}
Compared to the corresponding values (\ref{limits_opt}) of the optimal protocol, the constant control protocol (\ref{constant_u}) behaves worse for smaller values of $\alpha$, while both strategies behave similarly for large $\alpha$. This is also evident in Fig. \ref{fig:efficiency}, where the conversion efficiency of protocol (\ref{constant_u}) is displayed as a function of the optical distance (blue dashed line). 

As a specific example we consider the case where $\alpha=100$. Solving numerically transcendental Eq.\ (\ref{transcendental}), we obtain for the optimal protocol $\theta_0\approx 1.540568$ rad. We subsequently use this value in Eq. (\ref{u_theta}) and find $u_s\approx 0.015105$. The corresponding dimensionless spatially-dependent control fields are displayed in Fig. \ref{fig:controls} (red solid lines), as functions of the dimensionless distance $\zeta=z/L_{abs}$. In general, control field $\Omega_c$ precedes $\Omega_d$ in space, in order to prepare the necessary atomic coherences for the conversion $\Omega_p$ to $\Omega_s$, but note that for the optimal protocol $\Omega_d(0^+)$ and $\Omega_c(\alpha^-)$ have nonzero values, associated with the jumps in the mixing angle at $\zeta=0$ and $\zeta=\alpha$, respectively. The corresponding dimensionless intensities of the probe and signal fields are plotted in Fig. \ref{fig:prop_theory}, with red solid lines when using the approximate propagation equation (\ref{system}) and with green circles for the full Maxwell-Bloch set of equations (without the weak loss assumption $\rho_{11}\approx 1$). Observe the very good agreement between the two results. For comparison, we consider a pair of adiabatic controls that we have used in Refs.\ \cite{Hamedi19,Paspalakis02}, of the form
\begin{equation}
\label{adiabatic_controls}
\Omega_c=\Omega_0\left[1+e^{(\zeta-\zeta_0)/\bar{\zeta}}\right]^{-1/2},\Omega_d=\Omega_0\left[1+e^{-(\zeta-\zeta_0)/\bar{\zeta}}\right]^{-1/2}
\end{equation}
Here, we take $\zeta_0=\alpha/2=50$ and $\bar{\zeta}=5$, where the latter value is selected such that the adiabaticity condition  $|\dot{\theta}|\ll 1/2\Rightarrow\bar{\zeta}\gg 1/2$ is satisfied. The dimensionless control fields are plotted in Fig. \ref{fig:controls} (blue dashed lines) as functions of the dimensionless distance $\zeta=z/L_{abs}$, while the corresponding dimensionless probe and signal pulses in Fig. \ref{fig:prop_theory}, with blue dashed lines for the approximate propagation equation (\ref{system}) and with yellow squares for the full Maxwell-Bloch equations. As before, a very good agreement is observed between the two results. The conversion efficiency at the final distance $\zeta=\alpha=100$ for the adiabatic protocol (\ref{adiabatic_controls}) is 0.8197, much lower than the value 0.9094 obtained with the optimal protocol. The corresponding efficiency of the constant control protocol (\ref{constant_u}) is 0.9077, close to the optimal one as shown in Fig. \ref{fig:efficiency}. For efficiency values close to 1 one has to use larger values of $\alpha$, a few hundred for the optimal protocol and about 1500-2000 for the adiabatic protocol. Note that in our recent work \cite{Hamedi19}, using adiabatic controls in the direct configuration, we have reported a conversion efficiency close to unity for optical densities as small as $\alpha=40$, but recently discovered that this result was overstated due to a scaling error. 

In summary, we showed that for two frequently encountered configurations of the double-$\Lambda$ atom-light coupling scheme, with the control fields applied in the same and different $\Lambda$-subsystems, the forward propagation of the probe and signal fields obeys the same set of equations. We also used optimal control theory to find the optimal spatially-dependent control fields which maximize the conversion efficiency from the probe to the signal field, for a single-pass scheme and a specified optical density. Aside this ``optimal protocol" (\ref{pulse_sequence}) we also considered two other, called the ``constant control protocol'' (\ref{constant_u}) and the "adiabatic protocol" (\ref{adiabatic_controls}). This work can be used in the implementation of efficient frequency and orbital angular momentum conversion devices, but also in other applications involving the double-$\Lambda$ atom-light coupling scheme.

\noindent\textbf{Funding.} State Scholarships Foundation (MIS-5000432).

\noindent\textbf{Acknowledgements.} Financed by Greece and the European Union (European Social Fund-ESF) through the Operational Programme ``Human Resources
Development, Education and Lifelong Learning", project ``Strengthening Human Resources Research Potential via Doctorate Research" (MIS-5000432), implemented by the State Scholarships Foundation (IKY).





\end{document}